# In Search of a Taxonomy for Classifying Qualitative Spreadsheet Errors

Zbigniew Przasnyski, Linda Leon, and Kala Chand Seal
Loyola Marymount University, Los Angeles, California (USA)
zprzasny@lmu.edu

**ABSTRACT**

*Most organizations use large and complex spreadsheets that are embedded in their mission-critical processes and are used for decision-making purposes. Identification of the various types of errors that can be present in these spreadsheets is, therefore, an important control that organizations can use to govern their spreadsheets. In this paper, we propose a taxonomy for categorizing qualitative errors in spreadsheet models that offers a framework for evaluating the readiness of a spreadsheet model before it is released for use by others in the organization. The classification was developed based on types of qualitative errors identified in the literature and errors committed by end-users in developing a spreadsheet model for Panko's (1996) "Wall problem." Closer inspection of the errors reveals four logical groupings of the errors creating four categories of qualitative errors. The usability and limitations of the proposed taxonomy and areas for future extension are discussed.*

## 1. INTRODUCTION

The prevalence of errors in organizational spreadsheets that can lead to disastrous consequences is well documented, yet few organizations have formal policies for quality control of spreadsheets deployed for end-users (Caulkins et al., 2007; Panko and Ordway, 2005; http://www.eusprig.org/stories.htm). The literature on spreadsheet errors classifies errors into two general categories: quantitative and qualitative (e.g., Beaman et al., 2005; Panko & Aurigemma, 2010; Powell et al., 2008; Teo & Tan, 1999). Quantitative errors are identified as immediate incorrect numerical values or logic in the spreadsheet, while qualitative errors are associated with spreadsheet design flaws that increase the likelihood of an eventual quantitative error occurring during operational use of the spreadsheet. To fully assess the quality of a spreadsheet model and certify it for operational use, it is therefore necessary to identify the presence of both quantitative and qualitative errors in the model. In this paper we propose a taxonomy for classifying qualitative errors. The scarcity of formal taxonomies for classifying qualitative errors in the literature is the motivation for this paper. The paper first provides some background on the various types of spreadsheet errors and existing taxonomies followed by a detailed description of our own proposed taxonomy, and a discussion about its usability and the conclusion.

## 2. SPREADSHEET ERROR TAXONOMIES AND QUALITATIVE ERRORS

The taxonomy proposed by Panko and Halverson (1996) was the first significant attempt to create a classification of errors in spreadsheets. Since then a number of researchers have contributed to the development of spreadsheet error taxonomies, but most work has focused on quantitative errors (e.g., Panko & Sprague, 1998; Powell et al., 2008; Rajalingham et al., 2000). Very little work has been done so far on developing taxonomies to classify qualitative errors even though qualitative errors can be just as damaging to an organization's productivity as quantitative errors (Panko and Aurigemma, 2010). Panko and Halverson's original quantitative error taxonomy was subsequently revised (Panko and Aurigemma, 2010; Panko and Halverson, 2001) to include the effect of the spreadsheet life cycle on errors, thereby distinguishing development and design errors from errors generated by users during the operational use phase of the spreadsheet.

**Figure 1: Panko and Aurigemma (2010) spreadsheet error classification revisited**

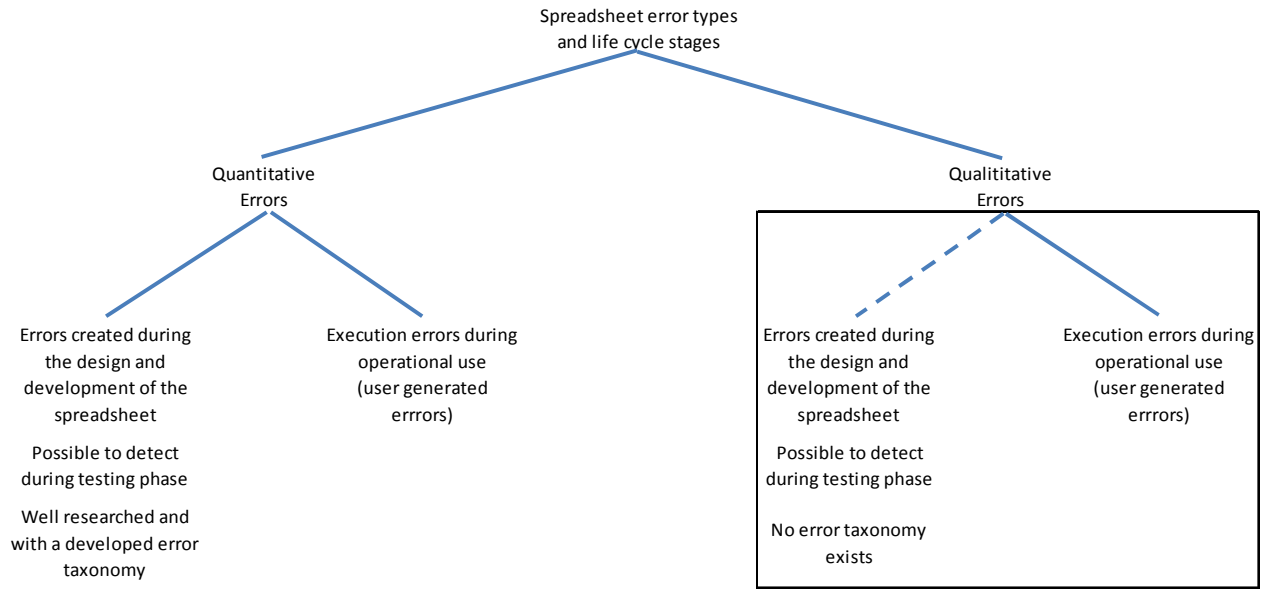

Panko and Aurigemma's classification is extended in the boxed area of Figure 1 to indicate that qualitative errors can be generated both during the design and development as well as operational use stages of the life cycle, just like quantitative errors. In this paper, we are interested in developing a taxonomy for qualitative errors created during the spreadsheet design and development phases so that they can be detected during the testing phase before the spreadsheet is released for operational use – indicated by the broken line in Figure 1. Qualitative errors, such as entering an inappropriate input value or overwriting a formula with a number (labeled as execution errors in Figure 1), may be generated by end-users during operational use of the spreadsheet, but we argue that avoidance of such errors is dependent upon the use of appropriate spreadsheet controls and user training. Some examples of controls that can be considered to reduce user-generated qualitative errors are cell protection, data validation, and conditional formatting. (Bewig, 2005; Martin, 2005; O'Beirne, 2005).

Design related qualitative errors are difficult to classify because the exact classification of the error is often based on its context, which is dependent upon the view of the reviewers. This same difficulty has been observed by researchers attempting to outline "best practices" for spreadsheet development (e.g. Conway & Ragsdale, 1997; Raffensberger, 2003). Examples of design related qualitative errors considered in past research include semantic errors, jamming or "hard-coding" errors, duplication errors, and poor design layouts that make the model hard to read (Purser & Chadwick, 2006; Raffensberger; 2003; Rajalingham et al., 2000; Teo & Tan, 1999). Clearly, identifying and avoiding these types of errors improves the quality of the spreadsheet. However, this list is far from complete and there is a need for systematic research to identify and classify these and other types of qualitative errors that can occur in spreadsheets. Panko and Aurigemma (2010) note: "It is now time to shift our focus into qualitative errors, which may be far more common than quantitative errors, and identifying the large number of different types of errors that are possible in different life cycle stages and by people with different roles to play" (p.244).

## 3. PROPOSED QUALITATIVE ERROR TAXONOMY

We describe the development of a qualitative error taxonomy that should help an organization detect if a spreadsheet contains certain types of qualitative errors, which potentially can be damaging if the spreadsheet is allowed to go into operational use. The purpose of the taxonomy is to help managers decide whether a numerically accurate spreadsheet (i.e., containing no quantitative errors) is robust enough to be in operational use. This would help tremendously in spreadsheet governance as auditors and testers of spreadsheets will have a mechanism for judging the spreadsheet quality for usability.

We took a two-pronged approach in developing the taxonomy. We first consolidated and organized the qualitative errors found in the literature using the umbrella taxonomy for qualitative errors outlined in Rajalingham et al., (2000). Second, we analyzed the qualitative errors found in a lab experiment that consisted of 104 spreadsheet models created for the "Wall Problem" described in Panko (1996). We found that the Rajalingham et al., (2000) qualitative error taxonomy was not sufficiently detailed in its practical use as a rubric for detecting and counting the various types of qualitative errors that were present in these completed spreadsheets. We therefore expanded the taxonomy by introducing specific elements that could measure the various types of qualitative errors at a more atomic level and would provide a set of rubric items for a formal grading process. Figure 2 expands the box in Figure 1 to provide a detailed picture of how our taxonomy fits in the overall classification scheme of qualitative spreadsheet errors.

**Figure 2: Proposed qualitative error taxonomy**

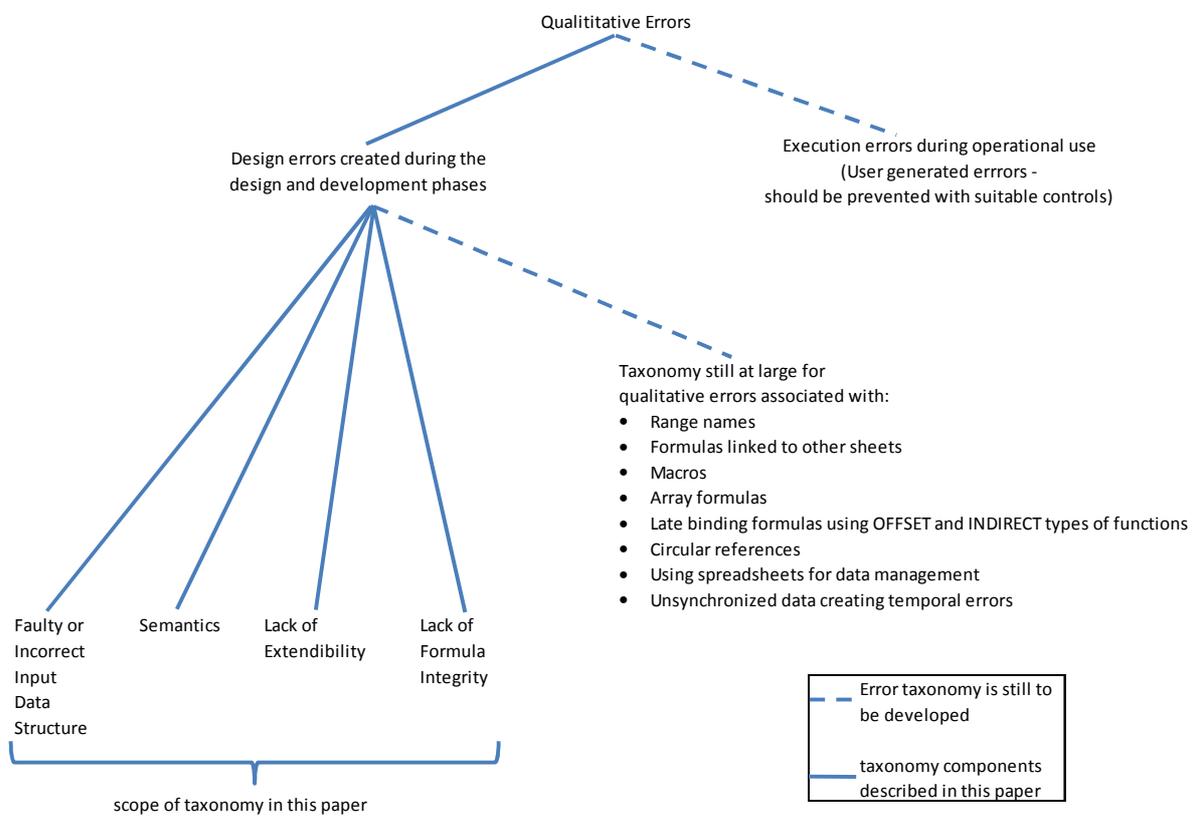

In the proposed taxonomy, as long as one instance of a particular type of qualitative error was present in the spreadsheet we recorded it as 1. Initially we tried to count every instance of a qualitative error but found the process to be inefficient and inconsistent across reviewers. Since qualitative errors are latent errors (Reason, 1990) that often lead to serious errors later, we argue that it is important to ensure that the presence of each type of error is flagged and recorded so that the spreadsheet model can be revisited before it is released for operational use. Depending on the domain, the mere existence of a certain type of error (e.g., hard-coding of the interest rate in a financial spreadsheet that updates based on a rolling interest rate) should trigger a complete reexamination of the spreadsheet for detection and correction of all instances of that error. The resulting list of the qualitative errors from our lab experiment and their subsequent logical groupings is shown in column 2 and column 1 of Table 1 respectively.

**Table 1. A taxonomy for defining qualitative errors**

| Category | Type of Error | Record the error as 1 ("exists") if: | Explanatory Example |
|---|---|---|---|
| Input Data Structure | Hard-coding/jamming values into formulas | At least one value is hard-coded into a formula somewhere in the model | 1-1 |
| | Duplication of Input Values | • There is more than one place to enter the same value in the model<br>• An intermediate calculated value must be entered directly as an input in another part of the model | 1-2 |
| | Input Cells Not Clearly Identified | One or more input cells anywhere in the spreadsheet are not clearly identified | 1-3 |
| Semantics | Missing Cell Documentation | At least one cell where a number is displayed is not labeled | 2-1 |
| | Incorrect Cell Documentation | At least one cell has a label that is clearly incorrect (e.g., refers to total profit when it is total cost; or states wrong units) | 2-1 |
| | Ambiguous Cell Documentation | At least one cell **that has not been identified** as clearly incorrect has an ambiguous label associated with it (e.g., no units are specified, general labels such as "costs" are used) | 2-1 |
| | Poor Layout for Readability | The spreadsheet is generally difficult to read/follow for any of the following reasons:<br>• Does not read left to right and top to bottom<br>• Related formulas are not in physical proximity of each other<br>• Excessive blank spaces are used to create long arcs of precedence<br>• The logic of the business problem is difficult to follow | 2-2 |
| Extendibility | Poor Layout for Model Extension | • Inserting a row or column will hurt the readability of the model or<br>• Extending the model will require multiple insertions and manipulation of the model | 3-1 |
| | Poor Layout for Copy/Paste | Layout of the logic of the model does not allow for easy copying to a new column/row | 3-2 |
| | Poor Absolute/Relative Cell References for Copy/Paste | At least one cell included in a layout that allows for the cell logic to be easily copied to another row or column contains a formula with incorrect use or lack of use of the $ notation | 3-3 |
| Formula Integrity | Spurious Formulas | A confusing or spurious entry in a formula exists | 4-1 |
| | Lack of Explicit Formulas | An inputted number (jammed or entered in an input cell) is the result of a calculation not carried out in the spreadsheet (even if some of the numbers used to calculate the input value were generated in the spreadsheet) | 4-2 |

**Category 1: Input Data Structure**

The Input Data Structure error category focuses on inappropriate structures used for incorporating input data values into the model's formulas and logic. This category includes the two qualitative errors most commonly referred to in existing spreadsheet literature, the hard-coding or "jamming" of a value in a formula and the duplication of an input value. In both cases, the assumed data value may initially be correctly entered thus creating no quantitative error. However, as a user interacts with the model and performs what-if analyses, the user may fail to correctly update all of the places where the input data value assumptions exist, and thereby generate an error in the bottom-line result.

*Example 1-1: Hard-coding/Jamming Values into Formulas*

Assume that three numbers representing the height, width and length of a wall are jammed into the volume formula in cell C3 of Table 2. Here the hard-coding hides the dimension inputs so that the user does not clearly see what assumptions were made for the current bottom-line output. The user may forget that the wall is assumed to be 6 ft. tall, 2 ft. wide and 20 ft. long since these inputs are not clearly documented. Furthermore, when the user wants to re-dimension the wall, the user must correctly identify which number, (6, 2 or 20), in cell C3 represents the dimension to be modified. There is a distinct risk that the wrong number will be modified.

**Table 2: Example of Hard-Coding**

| Values | | | | Formulas | | |
|---|---|---|---|---|---|---|
| | B | C | | | B | C |
| 2 | Type of Wall | Volume (in cubic ft) | | 2 | Type of Wall | Volume (in cubic ft) |
| 3 | Lava Rock | 240 | | 3 | Lava Rock | =6*2*20 |

Absolute constants, such as pi or the number 1 when it represents 100%, can be hard-coded without generating these risks. These are not jamming errors as these numbers cannot be changed in subsequent use and the functional purpose of the number in the formula is clear to the user.

*Example 1-2: Duplication of Input Values*

Assume that for the "Wall problem" the company requires a 30% profit margin on all of its bids. The required 30% profit margin has been jammed into both profit margin formulas in cell C14 and D14 of Table 3. This qualitative error is more serious than the simple hard-coding errors described in Example 1-1 as the user must change the profit margin in two places in order to generate the correct bottom-line result. There is a risk that the user will remember to change the profit margin correctly in only one of the two locations, forgetting about the other formula.

**Table 3: Example of Duplication**

| Values | | | | | Formulas | | | |
|---|---|---|---|---|---|---|---|---|
| | B | C | D | | | B | C | D |
| 11 | | Type of Wall | | | 11 | | Type of Wall | |
| 12 | | Lava | Brick | | 12 | | Lava | Brick |
| 13 | Total Costs | $1,296 | $1,056 | | 13 | Total Costs | =C10 | =D10 |
| 14 | Required Profit Margin | $388.80 | $316.80 | | 14 | Required Profit Margin | =0.3*C13 | =0.3*D13 |
| 15 | Bid | $1,684.80 | $1,372.80 | | 15 | Bid | =SUM(C13:C14) | =SUM(D13:D14) |

This model has two types of errors which should be recorded as such to reflect its higher risk of eventual misuse. Here two numbers have been hard-coded into formulas as in example 1-1 and the value for the same input must be entered in two different places, creating an additional duplication error. However, duplication errors may or may not occur simultaneously with instances of hard-coding. If, say, cells C5 and D5 in the Table 3 example had been set up as data input cells for the profit margin rate and correctly referenced in C14 and D14, then there is no hard-coding error, but the need to change the profit margin in two places (assuming that they are never to be different for each wall) creates a duplication error.

*Example 1-3: Input Cells Not Clearly Identified*
Input assumptions may be labeled clearly with appropriate cell documentation but still be difficult for users to identify if the model layout does not properly emphasize the location of the input cells. There are several ways to clearly identify input cells for the user. The traditional modular computer programming approach, which is commonly referenced in spreadsheet literature, specifies standard formats with a separate and clearly labeled data input module. Use of a designated color may alternatively be used to identify input cells that are not separated in a distinct module.

It should be noted that this error classification should not include hard-coding or duplication errors. This error type presupposes appropriate input data structures (e.g., no jamming) but identifies layout and formatting design flaws where the input cells may not be easily recognizable or understood by the user.

**Category 2: Semantics**

The Semantic error category describes errors that create a distortion of or an ambiguity in the meaning of an input or output of the model. As a result, the user can make an inappropriate choice of an input assumption value and/or a misinterpretation of an output result. Common sources of semantic errors are poor documentation of assumptions and outputs, including cell formatting and labeling, as well as the readability of the layout design.

*Example 2-1: Cell Documentation Problems*
*Cells which are not labeled* create confusion. Some of these cells may be a key calculation in the model's logic while others may be dangling cells that serve no purpose. In both cases, unlabeled cells add visual and logical clutter and confuse the user.

*Cells which are incorrectly labeled* or formatted often result in wrong inputs and/or decisions being made. For example, in Table 4 due to the missing documentation in row 13 and an incorrect label in row 15, a user might not realize that the required profit margin is already included in row 15 and might add a second mark-up to row 15's results to calculate the "final bids".

**Table 4: Examples of Missing and Incorrect Cell Documentation**

|    | B | C | D |
|----|---|---|---|
| 11 |   | Type of Wall | |
| 12 |   | Lava | Brick |
| 13 |   | $1,296 | $1,056 |
| 14 | Margin | $388.80 | $316.80 |
| 15 | Total Cost | $1,684.80 | $1,372.80 |

Misunderstandings arising from incorrect cell documentation can occur for inputs as well as outputs. For example, if the input cell containing the fringe benefit percentage was incorrectly formatted to display as a $ value, the user may convert the percentage input value into a dollar equivalent estimate, which would be an inappropriate unit for the model's logic.

*Cells with ambiguous documentation* can be just as misleading as missing and incorrect cell documentation errors. For example, in Table 5, it is not obvious that $20.00 in cell C2 represents the wage rate per hour or if the fringe benefit in cell C3 is a percentage of the wage rate as opposed to a dollar pay rate. In both cases, the user will have to make implicit assumptions about the context of the input when a value is entered for the cell, thus creating the possibility of error.

### Table 5: Examples of Ambiguous Cell Documentation

|   | B | C |
|---|---|---|
| 2 | Wages | $20.00 |
| 3 | Fringe Benefit | 0.2 |

Similarly, confusion regarding outputs can occur when the labeling or formatting is not as detailed or numerically as precise as it should be. Incorrect formatting can display a rounded value of a fraction (e.g., 0.80 displayed as 1) and thus create confusion or lead to acceptance of an incorrect result. Similarly, when a cell is labeled *Costs*, it may be interpreted as *Total Costs*, *Material Costs* or *Labor Costs*. Ambiguous cell documentation is one of the most frequent qualitative error and yet one of the hardest errors to detect. The reviewers in our study had more difficulties in consistently identifying ambiguous cell documentation errors than any other type of qualitative error. To maximize the likelihood of detecting these errors, testing needs be done by an independent party that is not familiar with the model who can be more objective about the confusion created by the ambiguity.

*Example 2-2: Poor Layout for Readability*
Conway & Ragsdale (1997) emphasize the importance of readability in spreadsheet design and present general guidelines to organize a spreadsheet for readability. They stipulate that the spreadsheet should be laid out so that the user is reading left to right and top to bottom, related formulas should be in physical proximity of each other, use of blank space should be carefully used to distinguish blocks but not create unnecessarily long arcs of precedence, and the model should be kept on one screen or worksheet if possible. Within these general guidelines however, there are still different ways that a developer may choose to organize and communicate the business logic of the problem being modeled (e.g. Raffensberger, 2003).

Some organizations have created standards that govern layouts and data input cell formats to promote consistent development practices that will make spreadsheets easier to review, test and use. If such standards exist within an organization then the evaluation for Examples 1-3 and 2-2 should be modified with respect to how well the model follows the company's specified guidelines. If standardized layout development practices do not exist in the organization, the general impression of the model's readability can be evaluated. In our study, existence of poor layout for readability error was recorded if the reviewer could not easily follow the business logic of the problem.

**Category 3: Extendibility**

The Extendibility error category focuses on qualitative errors that place the model's integrity at risk when the model is to be used as a template by others and may need to be extended to make it operational for its new application. A model designed for readability by the user in the current business application may not be designed for generalization to other similar applications or for extensions of the existing problem's dimensions, such as additional time periods or product lines. There are certain layout attributes that increase the usability of a model when basic dimension and line item modifications may be required.

*Example 3-1: Poor Layout for Model Extension*
Models must be laid out such that when a row or column is inserted to extend a particular section of the model, the insertion does not hurt the readability of the other sections. In addition, a layout should require as few insertions of rows/columns, copy commands and formula modifications as possible into the existing model sections for extending the model.

Table 6 presents a model that is considered very readable but whose readability would be compromised should the user want to include a third type of material in the bidding process. It is easy to copy the calculation logic in cells D20:E29 to a new section for the third material bid in cells G20:H29. The cost per cubic foot in row 21 however is linked back to cell B5 in the input section. To maintain the integrity and readability of the model, a row would need to be inserted after row 5 for the new material information. With the other input assumption sections laid out horizontally, the blank row will extend (with input shading for some cells) into these sections and may be confusing for future users. In addition, the user would have to know to link the material cost in the new cell H21 to the new input cell B6. Thus the model has a poor layout for extension because the user would need at least three steps to modify the model to include a third building material and the resulting model would be less readable after the modifications.

**Table 6: Example of Poor Layout for Model Extension**

|    | A | B | D | E | G | H |
|----|---|---|---|---|---|---|
| 1  | Assumptions | | | | | |
| 3  | Cost of Building Materials (per cu. ft) | | Wall Dimensions (Ft.) | | Labor Estimates (1 Wall) | |
| 4  | Lava Rock | $ 3.00 | Length | 20 | # Days | 3 |
| 5  | Brick | $ 2.00 | Height | 6 | # Hrs./Day | 8 |
| 6  | | | Width | 2 | # People in Crew | 2 |
| 7  | Bid Margin: | 30% | | | Fringe Benefit Percentage | 20% |
| 8  | | | Total Cubic Volume | 240 | Hourly Wage | $ 10.00 |
| 9  | | | | | | |
| 10 | | | | | Total Projected Hours (per wall) | 48 |
| 11 | | | | | | |
| 12 | | | | | Labor Salary Expense | $ 480.00 |
| 13 | | | | | Fringe Benefit Expense | $ 96.00 |
| 14 | | =Input Field | | | **Total Labor Expense** | **$ 576.00** |
| 15 | | | | | | |
| 18 | Calculations | | | | | |
| 20 | **Lava Rock** | | **Brick Rock** | | | |
| 21 | Lava Rock Cost/cu.ft. | $ 3.00 | Brick Rock Cost/cu.ft. | $ 2.00 | | |
| 22 | Cubic Feet of Wall | 240 | Cubic Feet of Wall | 240 | | |
| 23 | Total Material Costs | $ 720.00 | Total Material Costs | $ 480.00 | | |
| 24 | Labor Expense | $ 576.00 | Labor Expense | $ 576.00 | | |
| 25 | | | | | | |
| 26 | Total Expected Cost | $ 1,296.00 | Total Expected Cost | $1,056.00 | | |
| 27 | Bid Margin | 30% | Bid Margin | 30% | | |
| 28 | | | | | | |
| 29 | **Grand Total Bid** | **$ 1,684.80** | **Grand Total Bid** | **$1,372.80** | | |
| 30 | | | | | | |

*Example 3-2: Poor Layout for Copy/Paste*

Some models will be laid out so that rows/columns can be easily inserted without harming the readability of the model, but the layout of the logic in the model does not allow for easy copying to the new row/column. For example, in Table 7 the labels between inputs and calculations are switched. Columns D and E, that contain the cost inputs, are labeled as Lava and Brick wall types respectively, but columns F and G are labeled as Brick and Lava respectively, thereby not allowing to copy the relative formulas in column F to column G.

**Table 7: First Example of Poor Layout for Copy/Paste**

| A | B | C | D | E | F | G |
|---|---|---|---|---|---|---|
| Volume (in cu. ft.) | 240 | | Lava Cost/cu. ft. | Brick Cost/cu. ft. | Brick Wall Material Cost | Lava Wall Material Cost |
| | | | $3.00 | $2.00 | $480.00 | $720.00 |

Similarly, the labels for the cost inputs in Table 8 are placed in two consecutive rows (cells B11 and B12), but the headings for calculations are switched to columns (B15 and C15). This change in layout makes it impossible to copy the calculation logic from one cell to the other, as the relative reference relationship is lost. If another wall made of a third material is to be inserted in this spreadsheet, the user would have to insert a new row after row 12 and then manually program the logic into cell D15 as the formula in C15 cannot be extended through simple copying. This last step is considered a risky modification as it requires the user to identify correct references for the new formulas. Here, even though the spreadsheet was initially correct, a qualitative error exists in the form of the lack of extendibility through easy copy/paste and thus increases the chance of a quantitative error in subsequent use and modification of the spreadsheet.

**Table 8: Second Example of Poor Layout for Copy/Paste**

|    | A | B | C |
|----|---|---|---|
| 9  | Volume (in cu.ft.) | 240 | |
| 10 | | | |
| 11 | Lava cost/cubic ft | $3.00 | |
| 12 | Brick cost/cubic ft | $2.00 | |
| 13 | | | |
| 14 | | Lava Wall | Brick Wall |
| 15 | Total Material Cost | $720.00 | $480.00 |

Note that in evaluating the model layouts in Tables 7 and 8 for Example 3-2 (suitability for copy/paste), we did not evaluate whether the associated formulas had been correctly coded with the necessary absolute and relative references. The outcome in this example is based solely on how the various cells were positioned relative to each other.

*Example 3-3: Poor Absolute/Relative Cell References for Copy/Paste*

Even when the layout is designed well so that formulas can be copied and maintain appropriate relationships, the formulas may not be coded with the correct use of absolute and relative references for copying. For example the logic in cells D22:E29 in Table 6 can only be copied if references to B7, E8 and H14 are absolute and all other relationships are relative.

We only record this referencing error for formulas that could be copied **given the appropriate layout of the model** by assessing whether the correct use of absolute and relative references had been programmed to allow the formula to be successfully copied to other cells in the model. Based on this definition, no errors associated with Example 3-3 exist for the illustration in Table 8 as the model was not laid out to allow copying of the formulas. Determination of an error was based on the proper use of absolute or relative reference syntax in the formulas for various cells, specifically the incorrect use or absence of the $ symbol. These errors are not based on the accuracy of the existing cell formulas but rather are assessed under the assumption that a user will do a copy/paste in the future. The error occurs when the formulas in future pasted cells lose the integrity of the correct logic due to reference problems.

**Category 4: Formula Integrity**

The Formula Integrity error category considers how robust cell calculations are given the structure of the formula that was used to generate the cell entry. Assuming that the initial model was reasonably tested and no quantitative errors were identified, the logic and formulas associated with the cell should not compute a wrong result. However, some formula structures are vulnerable to misinterpretation and therefore are more prone to producing an eventual miscalculation once the model is put into operational use.

*Example 4-1: Spurious Formulas*

Formulas that use a function inappropriately create confusion about the correctness and intent of the formula. For example, if the formula *=sum(C13+C16)* is entered in a cell, then the use of the *sum* function is spurious and creates unnecessary confusion. While the intent of the formula is to add the entries in cells C13 and C16, a user may inadvertently read that formula incorrectly as *sum(C13:C16)*. This user would then believe that entries in cells C14 and C15 are being included in the total. A user who does notice that the parameter is indeed C13+C16 may wonder if it was supposed to have been C13:C16 and may be tempted to alter the formula to reflect a more common syntax. This will be especially confusing if additional line items are added between the current row 13 and row 16.

*Example 4-2: Lack of Explicit Formulas*

Models can be created where a value that has been calculated externally by the user is entered into an input cell, without the calculation or logic displayed in the spreadsheet itself. For example, if a formula did not exist for the wall volume in cell C3 of Table 2, the user externally calculated the volume of the wall and then input the resulting calculation in cell C3 of the model. This lack of explicit formula weakens the integrity of the model's results as there is no process to ensure that the user knows how to correctly calculate the volume of a wall and therefore may enter an erroneously calculated input into the model. This situation is even more serious if an externally calculated input is based on intermediate calculations generated in the model.

## 4. DISCUSSION

In developing the taxonomy presented in this paper, we referred to the guidelines suggested in Powell et al., (2008). They suggest that a well-developed classification should have three characteristics: (1) it should specify the purpose for which it was created and the context in which it is meant to be used, (2) each category should be clearly defined and examples provided, and (3) the classification should be tested in the relevant context and evidence provided that different people classify errors consistently. Our main context for investigating design qualitative errors was usability of the model by end users throughout the spreadsheet life cycle as described in the discussion of Figure 2. The taxonomy's ultimate purpose is to allow researchers and organizations to focus on identifying controls that can be used to mitigate the risks associated with each type of qualitative error associated with the design and development phase of spreadsheet models. We have also provided a clear definition for each type of error that can be detected during the testing phase of a spreadsheet model and provided examples of each type. The consistency of the taxonomy is not yet tested but an inter-rater reliability study is in progress.

Developing this taxonomy also has highlighted the issue of different approaches for counting the various qualitative errors. Unlike the more mature literature on quantitative errors where benchmarks for error counting are well established, the existing literature on qualitative errors does not yet have a consensus. For reasons of efficiency an organization may wish to adopt our methodology of simply flagging the existence of a type of error. It is also possible to count the number of cells that contain a particular type of error and calculate the cell error rate (CER) for each error type. In practice, an organization would need to define or adopt an error counting methodology that is appropriate for their purposes in evaluating the spreadsheet. However, for direct and meaningful comparisons outside the organization and for the research community at large, standardization of qualitative error counting is essential. We hope that our proposed taxonomy will be a starting point for researchers to develop and explicitly state a standardized qualitative error counting and classification methodology, so that future results can be meaningfully compared and measures can be recommended for reducing spreadsheet qualitative errors.

There are several limitations with the proposed taxonomy at present. The main limitation is that it is currently restricted to qualitative errors encountered in problems of a similar nature to the Wall problem. Possible other types of qualitative errors that result from use of features such as linking formulas between files, VBA, manual procedures for executing macros or copy/pasting data into the model, and use of array formulas and range names was not evaluated because these features are typically associated with modeling more complex problems. A classification of errors associated with use of these features could not be obtained from our experiment due to the simple nature of the "Wall" problem. The taxonomy should therefore be expanded and tested on problems that require larger models with more advanced features often embedded in an organization's large and mission-critical spreadsheets. Future research then would consider how the existing categories and subcategories in the proposed taxonomy could be extended and better defined so that more of a consensus emerges on all the possible types of qualitative errors.

## 5. CONCLUSION

Qualitative errors in spreadsheets are as serious as quantitative errors and their study is equally important. However, to date, there has been little discussion in the literature regarding qualitative errors. We have made an attempt to develop a qualitative error taxonomy based on the previous work in the literature and errors obtained from an experiment with the "Wall problem." We envision this research as a starting point for focusing on the definitions and counting schemes for the study of qualitative errors that can be used by the research community to move towards a broader, and generally accepted taxonomy for qualitative errors in a spreadsheet.